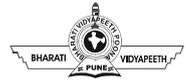

ORIGINAL RESEARCH

# Real-time monitoring as a supplementary security component of vigilantism in modern network environments

Victor R. Kebande[1] · Nickson M. Karie[2] · Richard A. Ikuesan[3]

**Abstract** The phenomenon of network vigilantism is autonomously attributed to how anomalies and obscure activities from adversaries can be tracked in real-time. Needless to say, in today's dynamic, virtualized, and complex network environments, it has become undeniably necessary for network administrators, analysts as well as engineers to practice network vigilantism, on traffic as well as other network events in real-time. The reason is to understand the exact security posture of an organization's network environment at any given time. This is driven by the fact that modern network environments do, not only present new opportunities to organizations but also a different set of new and complex cybersecurity challenges that need to be resolved daily. The growing size, scope, complexity, and volume of networked devices in our modern network environments also makes it hard even for the most experienced network administrators to independently provide the breadth and depth of knowledge needed to oversee or diagnose complex network problems. Besides, with the growing number of Cyber Security Threats (CSTs) in the world today, many organisations have been forced to change the way they plan, develop and implement cybersecurity strategies as a way to reinforce their ability to respond to cybersecurity incidents. This paper, therefore, examines the relevance of Real-Time Monitoring (RTM) as a supplementary security component of vigilantism in modern network environments, more especially for proper planning, preparedness, and mitigation in case of a cybersecurity incident. Additionally, this paper also investigates some of the key issues and challenges surrounding the implementation of RTM for security vigilantism in our modern network environments.

## 1 Introduction

The world has entered an era where individuals, organisations and government agencies want real-time faultless digital transactions and services on demand [1]. This also implies that any obstacle encountered while executing any of these transactions must be resolved in real-time. However, in the context of modern network environments, the challenges associated with on-demand real-time transactions and services are growing by the day as systems and networks become larger and complex [2]. This, therefore, makes it hard even for the most experienced network administrators to independently provide the breadth and depth of knowledge needed to oversee or diagnose the complex network problems that are prevalent. However, with good Real-Time Monitoring (RTM) knowledge network administrators can have constant and up to date information needed to make informed decisions immediately as well as see any trends as they unfold.

Note that in the context of this paper the authors define RTM as "the process of identifying or capturing data about faults, failures, performance, health and usage, slow systems among other issues that can cause organisation systems or network downtime and the continuous analysing and evaluating of captured data to help maintain, mitigate and optimize organisations' systems and network availability".

✉ Victor R. Kebande
victor.kebande@mau.se

[1] Department of Computer Science and Media Technology, Malmö Universitet, Nordenskiöldsgatan, Malmö, Sweden

[2] Faculty of Science, ECU - Security Research Institute, Edith Cowan University, Joondalup Campus, Australia

[3] Cybersecurity and Networking Department, School of Information Technology, Community College of Qatar, Doha, Qatar

**Table 1.** key vulnerable areas in MNEs and possible mitigation approaches

| | Network environment | Possible causative attacks | Possible security violations | Possible mitigation approaches |
|---|---|---|---|---|
| 1 | SCADA | Buffer overflow<br>SQL Injection [43] | Integrity<br>Availability<br>Confidentiality | Apply machine learning approaches to detect unusual behaviours through anomaly detection |
| 2 | Cyber-Physical Systems | Physical attacks<br>DDoS | Integrity<br>Availability<br>Confidentiality | Client–server monitoring/session monitoring. Monitoring solutions that query servers periodically |
| 3 | Critical Infrastructure Systems | DDoS<br>SQL Injection<br>Malware attacks | Availability<br>Integrity<br>Physical tampering | Robust and predictive monitoring and monitoring hyper logistics |
| 4 | IoT Ecosystems | Botnet attacks<br>MITM attacks<br>Identity theft<br>Advanced persistent threat<br>sensor-based attacks | Integrity<br>Availability<br>Confidentiality | Sensor instrumentation and monitoring CPS, exploit logging schemes |
| 5 | SDN Architectures | MITM<br>Sensor attacks<br>Social engineering | Integrity<br>Availability<br>Confidentiality | Logging/log analysis, network inventory, and parameter analysis of traffic. Acquiring and calculating network parameters in real-time |
| 6 | Cloud Infrastructures | DDoS<br>Botnet attacks | Integrity<br>Availability<br>Confidentiality<br>Authenticity<br>Privacy | Real-time acquisition of data to monitor traffic/telemonitoring systems |

From this definition, one can infer that increasing network visibility can help network administrators respond to network issues quickly, improves efficiency levels, and even give organisations the ability to spot possible system and network problems swiftly before they impact on key organisation transactions or services. This definition though does not make technological assumptions based on the modern network environment rather it has been inclined towards the technological determinants that cyberspace is modeled in.

As a contribution, this paper presents the importance of RTM as a supplementary security component of vigilantism in modern network environments. However, the paper also explores some of the key issues and challenges surrounding the implementation of RTM for vigilantism in our modern network environments. The objective is to show how RTM can help network administrators, analysts, and engineers respond to security incidents swiftly but also expose the issues and challenges surrounding RTM implementation.

The remaining part of this paper is structured as follows: Sect. 2 briefly presents the background followed by research advances in Sect. 3. Section 4 covers the relevance of RTM as a supplementary security component of cyber vigilantism in modern network environments followed by issues and challenges surrounding the implementation of RTM in Sect. 5. Key vulnerable RTM areas in modern networked environments are discussed in Sect. 6. Section 7 presents a critical evaluation of the propositions in this paper before a conclusion is made in Sect. 8 and make mention of the future research work.

## 2 Related Literature

This section presents a brief background of the following areas: the concept of a Modern Network Environment, Cyber Security Threats, and Real-Time Monitoring.

### 2.1 Modern network environment (MNE)

Traditionally, a network consists of two or more computers that are linked together to share resources, exchange files, or allow electronic communications [3, 4]. This definition though has evolved to accommodate new concepts that define the modern network environment as explained in this paper.

A Modern Network Environment (MNE) can thus be defined as a list of many interdependent components that include but not limited to servers, workstations, cloud infrastructure (cloud-based technology, Virtual Machines), instances, configurations, web applications, Supervisory Control and Data Acquisition (SCADA) systems, Software Defined Networking (SDN) architectures, Internet of Things (IoT) devices, the users and other network devices that need to be protected but at the same time allow network administrators have full visibility of their network. Also, MNEs are widely distributed with more bandwidth and buffering characteristics and may also include automation of some of the network activities as well as offer drag-and-drop interfaces. Besides, some of the modern network infrastructures are scalable, resilient, and always on. This makes MNEs have the ability to abstract away the need to know what vendor's equipment is being used hence simplifying the knowledge and expertise network administrators need to have.

MNEs also consider external services and remote users which sometimes need to be connected to the organizations' network daily. Because of their ability to access sensitive information and systems within the organization network, administrators need to have continuous monitoring into what these external services and users are doing, what devices they are accessing or using, what apps are they accessing and if they are using a virtual private network (VPN) and if at all they are following the laid down policies [5].

Despite all the technological developments, MNEs are still susceptible to cybersecurity threats. Some of the cybersecurity threats affecting MNE are briefly elaborate on in the next sub-section.

### 2.2 Cyber security threats affecting MNEs

A Cyber Security Threat (CST) is a malicious act by an adversary or a process that seeks to damage data, steal data, or disrupt digital systems in general [6].The outcomes of any CST may include but not limited to: theft of valuables, electrical blackouts, failure of military equipment, breaches of national security data, theft of sensitive medical data records, disruption of phone/mobile networks, interference with computer networks or paralyse computer systems making data unavailable and many other potential threats to human lives [7]. In MNEs, CSTs are relentlessly quirky, characterized by disguise and manipulation. They constantly evolve to find new ways to annoy, damage, or steal data, disrupt digital systems, or even harm individuals [8][10][11]. This, therefore, raises the need for network administrators to consider alternative ways of defending their networks. This, scenario motivated the research in this paper to consider RTM as a supplementary security component in modern network environments. The sub-sections to follow briefly explains some of the recent CSTs affecting MNEs.

#### 2.2.1 Crypto Jacking

With the distributed nature of our MNEs coupled with the fact that anything can be connected to something, Crypto-jacking has emerged as an online threat or malware that hides on a computer or mobile device and uses the machine's resources to "mine" cryptocurrencies [19]. Crypto-jacking can hijack web browsers, as well as compromise all kinds of devices, from desktops and laptops to smartphones and even network servers and IoT devices. Hackers use crypto-jacking to steal computing resources from their victims' devices to compete against sophisticated crypto mining operations without the costly overhead. According to [8, 9][10] this type of attack slows down computer processes, increases electricity bills as well as shortens the lifespan of computing devices. Hackers mostly use malicious links in an email that loads crypto mining code on the computer, or by infecting a website or online advert with JavaScript code that auto executes once loaded in the victim's browser.

#### 2.2.2 Ransomware

This is a form of malware targeting both human and technical weaknesses in MNEs to make critical data and/or systems inaccessible [12]. This type of malware is delivered through various vectors, including Remote Desktop Protocol, which allows computers to connect across a network, and phishing email [13] then spreads throughout the network by installing malicious software. In most cases, the hacker usually takes control of a computer or the networked device, locks its data, and demands a ransom from the victim promising to restore access to the data upon payment. When a compromised system is infected, the ransomware program shows a message requiring payment for functionality restoration. The payment is normally requested in bitcoin and is accompanied by the due date. The attackers use bitcoin transactions because it can anonymize the real culprit of the attack. Once the payment is done, in some cases, the system will be restored to normal. Unfortunately, it may not be 100% the case as some of the systems could be lost forever.

### 2.2.3 Cyber-physical attacks

According to [14], a cyber-physical attack on critical infrastructure occurs when a hacker gains access to a computer system that operates equipment in a manufacturing plant, oil pipeline, a refinery, an electric generating plant, or other similar infrastructure and can control the operations of that equipment to damage those assets or other property. Knowing that MNEs are dynamic and distributed, cyber-physical attacks are a risk not only for the owners and operators of critical infrastructures, but also for their suppliers, customers, businesses and persons in the vicinity of the attacked infrastructure, and any person or entity that may be adversely affected by it [14]. With cyber-physical attacks, a hacker can disable cameras, turn off a building's lights, make a car veer off the road, or a drone land in enemies' hands among many other attacks that can cause harm.

### 2.2.4 Endpoint attacks

Endpoints can be defined as points of access to an organisations network which create points of entry that can be exploited by malicious actors. This implies that an endpoint can be an IoT device, a local or remote computing device that communicates back and forth with a network to which it is connected to [15]. This may also include but not limited to Point-of-sale, industrial control or fixed-function devices, desktops, laptops, smartphones, tablets, servers, workstations among other devices. Endpoints in the context of MNEs are increasingly susceptible to cyberattacks thus making it easier for intruders to get past security measures. Endpoint security, however, is intended to protect these devices from malicious internal and external threats. Some of the biggest endpoint threats are phishing and ransomware attacks [16].

### 2.2.5 IoT attacks

Internet of Things (IoT) devices are a fast-growing sector of the Internet as well as in MNEs. However, as the number of connected devices proliferates, the increase creates additional points of access in a network. More access means more opportunity for unwanted entry points in the networks. This is true of any smart building, but it is especially worrisome in homes where everyone ought to feel safest [17]. It's fair to say that the proliferation and increased and continued ubiquity of IoT ecosystems has sparked a malicious attackers' renaissance, as most IoT devices require web access or mobile apps for manual control [18]. This access, in turn, can subvert some security developments of the past decade, exposing bugs that have been dormant [17].

With the growing number of IoT attacks in the world, many organisations have been forced to change the way they plan, develop, and implement cybersecurity strategies to reinforce their ability to respond to cybersecurity incidents. Hence, the need to consider RTM as a supplementary security component in MNEs. The next sub-section explains more about the concepts of R.

### 2.3 Real-time monitoring (RTM)

Real-time is a phrase mostly used to refer to the ability of a system to respond to something swiftly so that the response takes place almost at the same time as the event is happening [19]. For this reason, it has become very necessary to analyse real time data in many fields to enable quick reaction to momentary events [20]. In computer networks for example, if a switch port reports high bandwidth utilization, the network administrator will need to monitor the live utilization statistics of that port in real-time to know whether the problem still exists or not [21].

RTM, therefore, allows network administrators and engineers to see and respond to network events such as faults, failures, performance, health and usage, slow systems among other issues that can cause organisation systems or network downtime as they occur. The events data and information help them to review and determine the current status of network systems as well as the overall processes and activities executed on the network data in real-time, or as it happens. One key thing about RTM though is that for every command given by the network administrators the data or information returned is considered accurate as at the moment the command was issued and helps them make up to date informed decisions and see trends as they develop.

Some RTM tools can offer administrators with visual insights into the data which is fetched from various sources as well as instant notifications or alerts into specific data-driven or administrator-specified events, such as when some data values go out of range. Relevant data can then be displayed according to priorities and administrator preferences on customizable interfaces using graphs, bar graphs, pie charts, or percentages [20]. Research advances in RTM have opened new opportunities and applications in different fields, some of which are discussed in the section to follow.

## 3 Research advances in RTM

To begin with, [22] state that monitoring network statistics is important for maintenance and infrastructure planning for any network service provider. In this research, they showcase an initial analysis of a general-purpose network monitoring platform for high-speed mobile networks. They then use their developed platform as the basis for performing complex real-time analysis such as application usage behaviour, security analysis, infrastructure planning [22]. This research was well articulated however, it did not focus on the importance of RTM as a supplementary security component of vigilantism in MNEs as is the case of this paper.

In another research by Sultana and Jairam [23] argue that networks are growing extensively and managing huge networks is utterly challenging. Besides, they state that monitoring networks form an important part of network management which assists in visualization of the network behaviour in real-time. For this reason, they present in their study a few network monitoring approaches as well as various applications of network monitoring. Their approach, however, was also not geared towards the importance of RTM as a supplementary security component of vigilantism in MNEs [24] also highlighted in their study that, monitoring is an important concept in network management as it helps network operators to determine the behaviour of a network and the status of its components. They also state that Software-defined networking (SDN) is becoming increasingly popular for network provision and management tasks. It is on these grounds that they survey the tasks and challenges associated with SDN, providing an overview of SDN monitoring developments. Their research was purely based on SDN and not on RTM as a supplementary security component in MNEs.

More research by [25] proposes the development and implementation of better network monitoring solutions using SDN technologies. This was backed up by the fact that, as the complexity of the networks grows, the requirements for better network monitoring increase, and the traditional tools for network monitoring cannot meet these new challenges.

Another effort by [26] proposed an SDN-Monitor, which carefully selects switches to monitor to reduce resource consumption. Their research assumed that, with growing services running in clouds, it is critical to defending the services from Distributed Denial of Service (DDoS) attacks. For this reason, they argue that SDN provides a flexible platform for network monitoring and relies on a central controller to ask switches for traffic statistics to get global traffic visibility for security.

There also exist other research works on RTM, however, neither those nor the cited references in this paper have presented insights on RTM as a supplementary security component of vigilantism in MNEs in the way that is discussed in this paper. However, we acknowledge the fact that the previous research works on RTM have offered useful insights toward the presentation in this paper. In the section that follows, we explain the importance or relevance of RTM as a supplementary security component of vigilantism in MNEs which forms the first part of the contribution in this paper.

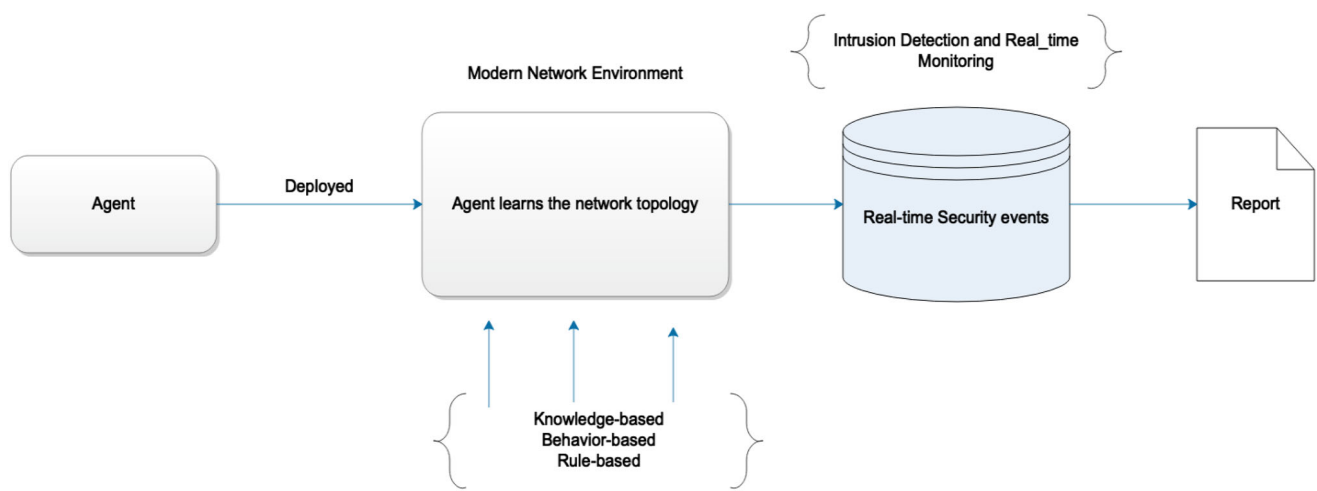

**Fig. 1** Agent in Modern learning Environment

# 4 Relevance of RTM as a supplementary security component of cyber vigilantism in MNEs

RTM has slowly become a key component of many organisations' networks. To understand the importance or relevance of RTM as a supplementary security component of cyber vigilantism in MNEs, this section digs deeper into revealing what value can an organisation obtain by considering RTM in its MNEs. Note that the list presented in this section has only what the authors randomly selected from documents and Literature surveys and deemed relevant to this study and does not in any way form a complete and comprehensive list. To begin with, Fig. 1 shows the need to have a real-time agent in modern network environments so that it can learn the changing nature of networks, which in turn can help in real-time monitoring.

Figure 1 shows a concept of a real-time agent deployed in a modern network environment that helps to learn the complexity of the environment concerning knowledge-based, behaviour-based and rule-based approaches and forward all that which is learned or the security events in a database based on real-time monitoring. An analysis of the data stored in the database can easily be presented in the form of reports. From this concept, many important and relevant features can be extracted that present new opportunities of interest to different organisations hence contributing to the relevance of RTM in the modern network environment. The list in the sub-sections to follow briefly explain the identified areas of relevance to this study with respect to RTM.

## 4.1 Centralized view of the network infrastructure

With the help of RTM tools in MNEs, irrespective of the networking devices and vendors, the underlying network infrastructure can be abstracted from many applications hence allowing the network administrators to have a single centralized view of everything that is happening in the network. Too many devices and applications without a centralized view can create problems in ways both big and small which could end up with network outages that cannot be traced quickly enough, cybersecurity risks, or even other compliance issues. However, when administrators can visualize the entire network in real-time from the port level to applications, then they can create significant savings in money, time, and other inconveniences as well.

## 4.2 Enhance efficiency: saving on time and resources

Traditionally some organisations use an old strategy that involves "run until it breaks" [27]. This strategy is based on reactive maintenance and only addressing things or events that are causing an active disruption into the network. Besides, this strategy can increase the cost exponentially. RTM, on the other hand, can help save the time needed by network administrators to manually inspect systems and devices for faults or noncompliance by automating the whole process which also reduces the incidence of human error thus lower costs. Setting priorities can also ensure that critical areas with high risks in MNEs can be addressed before they create more high-cost problems in the network.

## 4.3 Increased productivity

RTM can help keep an eye on all activities taking place in an organisations network environment through automation hence serving to prevent unauthorized transactions or services by identifying them quickly and responding to them swiftly. This implementation in MNEs, therefore, minimises the risks of network downtime giving the network administrators more time to complete other assigned tasks on time hence increasing productivity in the organisation. For people working from remote locations, they can get more work done with access to real-time information than when they must make un-ending calls to the organisation which also takes a lot of time to resolve some simple tasks.

## 4.4 Quick identification and detection of potential security vulnerabilities and threats

Effectively implementation and utilization of RTM can help identify and detect potential security vulnerabilities and threats. This is because RTM can quickly and efficiently help locate most of the problem's source, correlate data, and enable the organisation to swiftly mitigate a problem. Security breaches can cost an organisation a lot of money, but with good RTM one can improve reliability and lower overall support and ongoing maintenance costs. Vulnerability assessment and threat detection in MNEs is one of the greatest reasons for implementing RTM in many organisations.

## 4.5 Strengthen security controls

MNEs are larger and complex with virtualization being introduced in some organisation. Each new component added to the network means more security controls are needed to be implemented and tracked. Manually tracking or monitoring and analysing security controls can be both inefficient and inaccurate. With today's complex and hybrid network environments, having RTM in place can help strengthen security controls, keep up with organisations risk monitoring and compliance audits. This also means that organizations can streamline auditing and quickly respond to unauthorized transactions with ease. With RTM one can also integrate detective controls, preventive controls as well as reactive controls. According to [28], detective controls to monitor and compare access and permissions against actual network usage to ensure that monitored activities align with the authorizations granted. Preventive controls, on the other hand, examine policies and peer groups to determine if a problem could potentially arise. If an inconsistency is detected, then reactive controls take over and send alerts to notify the appropriate administrators so the issue can be corrected in real-time, before it becomes a larger risk to the organization.

### 4.6 Prevent data loss and data leakage

Data loss refers to an event in which important organisation data is lost while Data Loss Prevention (DLP) focuses on preventing unauthorized access, misuse, and illicit transfer of data outside organizational boundaries [29] DLP offers insight into where data lives and appropriately secures it besides showing real-time controls to detect violations of the data.

With DLP, RTM can be used to monitor organisations data and send an alert about attacker attempts to access any sensitive data as well as to detect and correlate network events that might constitute data leakage. This may also include automatically blocking any actions that violate the organisations security policies, data encryption and other protective actions to prevent end-users from accidentally or maliciously sharing data that could put the organization at risk [30]. Besides, it should also be possible to monitor access to all sensitive files and recording granular usage data such as user, department, file accessed, file type, and operation response time [29].

### 4.7 Optimize network performance

Network optimization is the process of making the best or most effective use of network resources in such a way that the network achieves maximum productivity with minimum wasted effort or expense for a given environment. Good implementation of RTM can give network administrators valuable insights into the organisation network's operation thus help them to manage bandwidth utilization, minimize latency, packet loss, congestion, and jitter. Optimizing network performance in MNEs can help observe the network performance and then create customized traffic models which in turn can improve the overall productivity.

### 4.8 Track network services functionality

With so many network services running on countless devices with different features and capabilities in MNEs, it can be overwhelming to manually monitor all of them. However, with the help of RTM, network services offered using different protocols such as SMTP, POP3, HTTP, TCP/IP, FTP among others can be monitored with ease thus giving the network administrator an easy way to track every single network service functionality.

### 4.9 Continually improve operations

With modern, dynamic, and complex network environments coupled with the increased cybersecurity threats, continually improving network operations is usually an effort of most network administrators. This helps to improve the functionalities and services offered by the organisations' network and deliver efficiency, effectiveness, and flexibility. However, this can also improve network quality, reduced network maintenance costs, simplified work processes, less waste of network resources, and improved service delivery. RTM can thus be used to monitor network operation and performance remotely and identify violations in real-time for fast remediation, which include but not limited to network security update services.

### 4.10 Understand the impact of any new changes

Whenever a new thing is introduced into the organisation network, there is always bound to be an impact. With real-time data analysis, administrators do not only get a complete picture of all the activities in a network but also are alerted of any changes or developments as they unfold. This helps them to address any problem promptly. Whichever the scenario, understanding the impact of any new changes in MNEs enables network administrators to quickly and accurately respond to critical issues or requests such as those that impact areas touching on healthcare data, automotive, and aerospace industry.

### 4.11 Monitor compliance and regulations

It is now becoming almost a must for all organisations to apply the requirements of compliance regulations and standards in all areas of business including the implementation of the Information Security Management System (ISMS) Framework. However, with many industries and government regulations coming into play, regulatory compliance obligations are becoming a challenge to many organisations, even more, challenging to those organisations operating in multiple geographic locations and with lots of staff members. For this reason, a simplified way to help organisation systems and networks adhere to regulatory compliance obligations could be the use of RTM and reporting tools to alerts organisations in real-time of any compliance policy violations. RTM can also be used to monitor certain issues such as network security and log management requirements imposed by many statutory and regulatory auditing authorities hence avoiding any compliance breach.

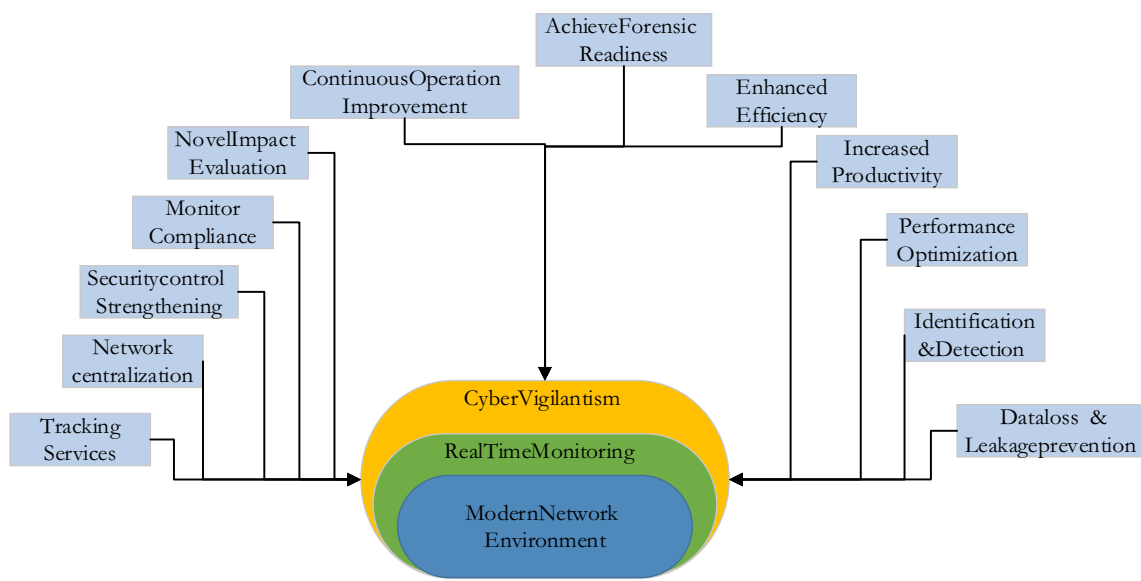

**Fig. 2** Overview of the Relevance of RTM in NME

### 4.12 Achieve digital forensic readiness

The readiness of an organization towards addressing digital forensics is a major drive in this current technology driven organizational workspace. Accordingly, studies [31, 32], have examined workplace readiness as the ability of an organisation to maximise its potential to use digital evidence whilst minimising the costs of an investigation. Good RTM of network events and gathering of potential evidence for forensic purposes can make the digital forensic investigation process much easier and faster. This is because digital forensic readiness allows an organisation to be properly prepared to handle security incidents and avoid damage and unnecessary costs [33, 34]. It is now becoming important to have forensic readiness in MNEs. Having looked at the relevance of RTM, the section that follows discuss the issues and challenges surrounding the implementation of RTM in any given MNEs. A depictive summary of the relevance of RTM as a potentially complementary component in NME is further presented in Fig. 2.

## 5 Issues and challenges surrounding the implementation of RTM

As good as the idea of having RTM implemented in one's organisation may sound, it has its own issues and challenges that organisations must battle with. Some of these challenges are briefly explained in the sub-section to follow.

### 5.1 Cost of hardware and software

Knowing that MNEs are dynamic, distributed, and complex, implementing RTM can be costly especially for small organisations. The hardware or software requirements may be prohibitively beyond reach to many small organisations. Besides, the maintenance and upgrading costs may also be too high for some organisations. With existing network environments integration may also not be simple hence adding up to the cost factor challenge to a different organisation.

### 5.2 Systems integration challenges

If all organisation were to use the same infrastructure and architecture in the world then systems integration would not be so much of an issue or a challenge. Unfortunately, this is not the case for many organisations in the world. Introducing new tools and technology in MNEs must be done with caution to avoid corrupting existing network environments and ensure future changes are taken into consideration. Different network topologies and architecture including different device vendors pause a huge integration challenge in MNEs.

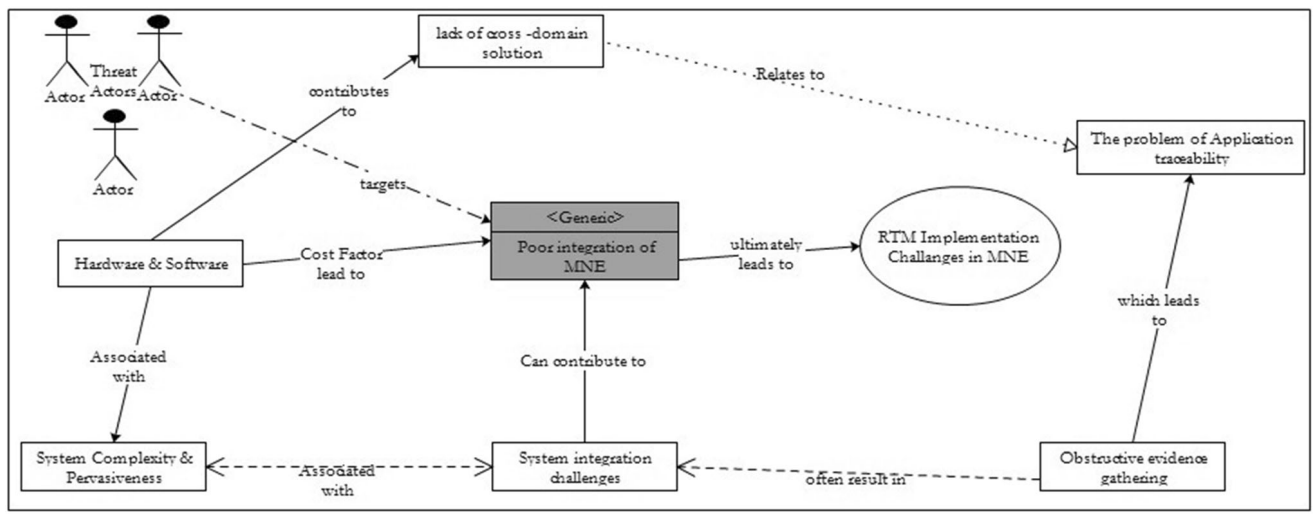

**Fig. 3** Perspective of Threat actors and the semantic composition of Challenges

### 5.3 The traceability of an application challenge

Traceability of an application in MNEs can be very tricky as it involves the capability to trace an application running in your network, its history and its location, and the ability to chronologically interrelate uniquely identifiable entities related to the application in a way that everything is verifiable [35]. A lot of research needs to be done in this area as this is one of the open challenges and problems in MNEs.

### 5.4 Lack of a cross-domain solution challenge

Complex interconnected systems are now present in MNEs. This, therefore, means that any organisation seeking to update or expand data and information-sharing capabilities in their networks must be able to do so without introducing security vulnerabilities to network. However, this is a big challenge as many vendors do not consider cross-domain solutions during hardware manufacture or software development. In an ideal situation Cross Domain Solution (CDS) technologies are meant to enable organisations to share data and information across physically, logically, and administratively separated networks in a reliable, secure, and interoperable manner. However, CDS creates a big challenge as different networks may have different security policies to address their exposure to different types of threats and levels of risk, and therefore hold differing levels of trust [36].

### 5.5 Summary of challenges to RTM implementation

The discussed challenges to the implementation of an RTM in a typical MNE is further summarized in Fig. 1. The RTM can indeed provide a baseline for the mitigation of the potential threats to the MNE. However, the diverse threat actors within the threat landscape tend to attack the process of MNE integration. Given that the effectiveness of a typical RTM is dependent on the integration-capability of the environment, potential threat actors are, therefore, capable of limiting the implementation of the RTM. This is further illustrated in Fig. 3.

One common challenge to the RTM implementation is the semantics between the individual challenges. For instance, the challenge of the lack of traceability of application can be attributed to two other challenges (obstructive evidence gathering, and the lack of cross-domain solution). These dependent challenges are also the consequence of other challenges as indicated in Fig. 3. This, therefore, implies that an attempt to provide an effective RTM within the current architecture of most MNE would require a redesign of the architectural composition of the MNE. Besides, the inherent configurational limitation of most devices within the MNE can further induce system challenges.

## 6 Key vulnerable RTM areas in modern networked environment

This section shows the key vulnerable RTM areas in MNEs that could cause attacks and respective security violations. However, possible approaches that could be used to enforce RTM are also discussed as is shown in Table 1. The key areas that have been considered represent the vulnerability brought by interconnected modern infrastructures that are susceptible to random attacks. Given the topology of the network and the ever-changing infrastructure, it is the authors' opinion that key areas that are affected by these attacks could easily be mitigated. In SCADA, for example, the authors have identified buffer overflow and SQL injection attacks that violate integrity (tampering with acquired data), availability, and confidentiality. To mitigate these kinds of attacks using RTM, some machine learning techniques that can statistically analyse these data to detect anomalies can be deployed. Others include Cyber-physical systems that can suffer from physical and DDoS attacks thus violating Confidentiality, Integrity, and Availability (CIA) triad. To mitigate these causative attacks, frequent monitoring of client–server sessions should be enabled. Another important area that is vulnerable is the critical infrastructure systems that can be affected by DDoS, SQL injection, and other malware attacks that violate availability and integrity. Based on these attacks it is important to enforce robust and predictive monitoring and hyper logistics [37].

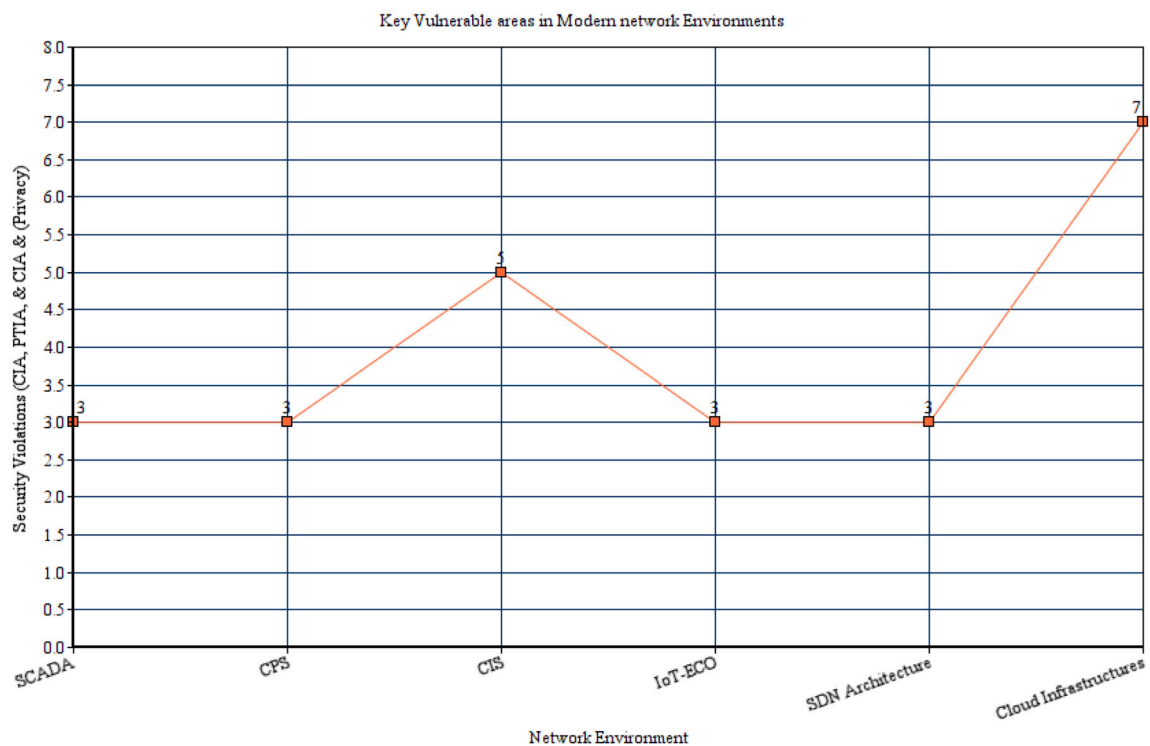

**Fig. 4** Severity of Key vulnerable areas of MNE

Next, are IoT attacks, which have in the recent past become more prevalent and are affected by botnet attacks, Man in The Middle attacks (MITM), identity theft, and Advanced Persistent Threats (APT). The security violation in this context is geared towards the CIA triad. Approaches to counter this include sensor instrumentation, monitoring, and enforcing logging mechanisms.

Other relevant vulnerable environment includes Software Defined Networking (SDN) architectures and cloud infrastructures which normally are affected by MITM, sensor attacks and social engineering aspects while the cloud can be affected by DDoS and botnet attacks. While SDN architectures and the cloud violates the CIA triad, the cloud infrastructures can further violate privacy because of the movement of data. Possible RTM approaches for SDN architectures include Logging/log analysis, network inventory, and parameter analysis of traffic [38–40]. Acquiring and calculating network parameters in real-time [41, 42] while for the cloud can include real-time acquisition of data to monitor traffic/telemonitoring systems. A summary of the key vulnerable RTM areas in modern networked environments is presented in Table 1 below.

Based on the key aspects that have been highlighted in Table 1, it is important to note that the mitigation approaches that the authors have stated are based on the possible generic approaches that have been used to defend the existing networked environment. While the authors do not beforehand suggest that only these techniques should be applied for all the malicious raids, it is imperative to explicitly say that novel approaches for the ever-changing environment could be used based on the prevailing scenario. Based on the vulnerable areas in MNE, we rank the network environments based on the severity (0–7) of the security violations as is shown in Fig. 4, where 7 is the most severe and 0 is the least severe. In this context SCADA, CPS, IoT ECO and SDN architectures are ranked as least severe with 3 because they violates CIA respectively. Next, CIS is ranked 5 which is mild severity for it violates integrity, physical tampering and Availability while cloud infrastructures are ranked as 7 because they violate CIAA and privacy.

Having looked at the key vulnerable areas, the next section provides a contextual critical evaluation.

## 7 Critical evaluation of the propositions

A discussion of the suggestions and propositions that have been put forward in this paper is given from a contextual approach in this section. To effectively monitor MNEs, it is paramount to identify key indicators and security events that are contained in the network infrastructure. Of importance in the MNEs is the schema that can be used in the analysis of security events that may affect the operation or the normal running of the networked environment. This, in the long run, may hinder the vigilantism that may help to avert adversarial behaviour. RTM involves, secure monitoring and logging of events, even though sometimes this may be done remotely in some instances. It is important when a correlation between what is conducted remotely and in real-time is mapped in the most effective way possible through dynamic reporting as was previously shown in Fig. 1. Consequently, numerous activities could be monitored and some of these activities may be denoted as key indicators or beacons that if attacked by adversaries may hinder the full operation of the network infrastructure. Also, it is important to note that important infrastructure supporting cyber-physical systems like SCADA networks may need real-time agents to be deployed so that they can learn the environment in an intrusion detection approach which utilises machine learning approaches to detect more sophisticated attacks.

In the context of this research, the key aspects that need to be monitored range from anomaly detection, real-time threat detection, and other attacks like integrity attacks and sensor-based attacks as is highlighted by [18]. That notwithstanding, most of the current networks have become indispensable and very open to quite a large and increasing number of adversaries. It is imperative to also note that the amount of data that in the recent past is being generated in these networks is massive and this is one of the important factors that necessitate RTM.

Because it is easy for adversaries to camouflage themselves while data is being exchanged, more resilient Intrusion Detection Systems (IDS) should be developed that can address the aforementioned painstaking challenges. The point of concentration in the past has always been to gain physical access to router and firewall logs. However, computing has shifted currently to the edge, and data is handled in a fog-based approach. Mostly, it becomes useful to access these data in real-time as they move across the edge hence the need for Adhoc networking infrastructure to monitor the data while profiling adversaries at the same time. it is also important to check the integrity, confidentiality, and privacy of these data while conducting real-time monitoring. The authors would like to emphasize that a key solution of uncovering and ensuring that the causative attacks that have been mentioned in Table 1 are dealt with, is to employ advanced anomaly detection techniques because the complexity of these attacks changes overnight rendering the network architectures to be exposed to adversaries. Modern advanced machine learning approaches can help one detect malicious activities with some degree of accuracy. Another important aspect worth exploring is network traffic, which at the time of writing this paper is still a contentious issue.

Contemporary techniques of analysing traffic in real-time have not addressed key challenges, i.e. Real-time forensic analysis of data has been proposed as a technique of conducting forensic logging and extracting digital data to see if a potential security incident could be detected in real-time [44–48]. Regardless of this, a suggestion by the authors is to adopt approaches that are more intelligent, knowledge-based and takes care of the behaviour of the network environment as shown by the agent-based approach in Fig. 1. By deploying such an agent, it could easily map the analysed traffic to possible vulnerabilities, while taking into consideration the amount of traffic at the disposal. The advantage of employing the approach shown in Fig. 1 is that the degree of identifying specific or known attacks, when an incident is detected may be used to develop approaches that exhibit some behaviour, which in the long run is a step toward attack detection.

Because of the foregoing, the authors posit that it is imperative to enforce RTM as a supplementary approach, and identifying the key security events in a network environment should be a point of concern. The data is the key player coupled with robust and more current attack techniques. While this paper has tried to present the networking, environment based on modern cross-cutting and more proliferated environments, it is also important to emphasize that RTM may well be enforced with agents that can interact and learn about the environment. Knowing that the cyber-space has become complicated and not easy to regulate, monitoring it needs real-time approaches. Therefore, if agents are trained to execute this task in real-time to understand the topology of the network infrastructure, then cyber-vigilantism could to some degree be enforced. Whilst diverse variations of RTM is currently implemented in most endpoint intrusion detection and response (EDR) system, the logic of an intelligent cyber-vigilantism presents a next-generation capability for the existing EDRs. Thus, the Authors would suggest a study in this direction, as a potential to harness the capability presented by the EDR and the intelligent cyber-vigilantism. Such a future direction should, however, not be limited to only EDR, as diverse RTM could leverage this integration.

## 8 Conclusion and future work

To this end, in addition to firewalls and virus scanners, other measures such as encoding software, data security software, content filters, port scanners, and other tools organisations need a good and comprehensive RTM approach. This paper highlighted the importance of Real-Time Monitoring as a Supplementary Security Component of vigilantism in Modern Network Environments. This, therefore, means that to enhance security in MNEs, RTM can play an important part in enhancing modern security measures. However, this paper also investigates some of the different key issues and challenges surrounding the implementation of RTM in MNEs and the key vulnerable areas. It is in the authors' opinion that this information will trigger more discussion and research into the field of Real-Time Monitoring in Modern Network Environments.

**Funding** Open access funding provided by Malmö University.